\documentclass[twocolumn,pra,aps,superscriptaddress,showpacs]{revtex4}

\usepackage{epsfig}
\usepackage{graphicx}
\usepackage{amsmath}

\newcommand{\beq}{\begin{eqnarray}}
\newcommand{\eeq}{\end{eqnarray}}
\begin{document}
\title{Controlling the Population Imbalance of a Bose-Einstein Condensate by a
Symmetry-Breaking Driving Field}
\author{Luis Morales-Molina}
\affiliation{Department of Physics and Center for Computational
Science and Engineering, National University of Singapore,117542,
Republic of Singapore}
\author{Jiangbin Gong}
\affiliation{Department of Physics and Center for Computational
Science and Engineering, National University of Singapore,117542,
Republic of Singapore} \affiliation{NUS Graduate School for
Integrative Sciences and Engineering, Singapore
 117597, Republic of Singapore}
\begin{abstract}
 Nonlinear Floquet states associated with a symmetry-breaking
driving field are exploited to control the dynamics of a
Bose-Einstein condensate in a double-well potential.  The population
imbalance between the two wells is shown to be controllable by
slowly tuning system parameters along a closed path. The results
extend symmetry-breaking-based quantum control to many-body systems
and extend navigation of linear Floquet states to navigation of
nonlinear Floquet states.
\end{abstract}
\date{\today}
 \maketitle

Symmetry-breaking driving fields have found a variety of
applications in quantum control, including the generation of
photo-currents by harmonic mixing \cite{kurizki}, the realization of
basic mechanisms of ratchet transport \cite{flach}, and the control
of chiral processes \cite{brumer}, to name a few.  To explore how
symmetry-breaking driving fields can be used to manipulate the
dynamics of Bose-Einstein condensates (BEC), it is necessary to
extend symmetry-breaking-based quantum control scenarios to
interacting many-body systems.

Given the robustness inherent in adiabatic quantum control, one
wonders how a driven BEC system may be manipulated by slowly tuning
the system parameters, especially those of the driving fields. To
that end one also needs to extend the conventional Floquet states to
mean-field nonlinear Floquet states
\cite{holthaus0,holthaus1,nonfloquet,njp}, a new concept for
understanding periodically driven nonlinear systems. Nonlinear Floquet
states  may display degeneracy as well as bifurcation points that
are absent in linear systems \cite{nonfloquet,njp}.  This presents
both challenges and opportunities for adiabatic quantum control. In
particular, the number of nonlinear Floquet states can differ from
the dimension of the Hilbert space, and that the adiabatic following
of nonlinear systems might necessarily break down due to
bifurcations \cite{Wu}.

In this study we demonstrate that the population imbalance of a BEC
in a double-well potential
\cite{kladko,smerzi,ober,double-well,kierig} can be controlled
(including its inversion) by using a symmetry-breaking driving
field. Specifically, by considering  a {\it closed} path of the
system parameters,  we show that it is possible to move one
self-trapped state from one well to the other, thus inducing an
inversion of the population imbalance and placing the system in a
different nonlinear Floquet state. 
Recognizing that a double-well model is relevant to other contexts
as well, e.g.,  a BEC occupying two hyperfine levels \cite{Rb} or a
BEC in an optical lattice that occupies two bands \cite{two-band},
this work shall motivate further studies of robust control of
many-body quantum systems with
symmetry-breaking driving fields.  

A BEC in a double-well potential can be readily realized
\cite{ober,double-well,kierig}. For example, the double-well
potential can be implemented by using two optical lattice potentials
\cite{bloch}. The periodic driving force can be created by
modulating the relative phase between the two optical lattices
\cite{kierig} (see also \cite{weitz}). Theoretically,  under the
mean-field two-mode treatment \cite{milburn}, the associated driven
dynamics can be described as follows:
\begin{eqnarray}\label{Eq:dimer1}
i \frac{d \psi_{1}}{dt}=C \psi_{2}+g\,\psi_{1} N _{1} +\psi_{1}
f(t),
\\
i \frac{d \psi_{2}}{dt}=C \psi_{1}+g\,\psi_{2} N_{2}-\psi_{2}
f(t),\label{Eq:dimer2}
\end{eqnarray}
 where $C$ denotes the tunneling rate constant, $g$ describes the
 nonlinearity due to the
 self-interaction of the BEC,
$N_{1,2}=|\psi_{1,2}|^2$ are the relative populations in the left or
right wells (under the normalization $N_{1}+N_{2}=1 $),
\begin{eqnarray}
 f(t)=f_{1}\sin (\omega t)+f_{2}\sin (2\omega t+\theta)
\end{eqnarray}
is a {\it zero-bias} bichromatic driving force with period
 $T=2\pi/\omega$, and $\theta$ is the relative phase between the $\omega$ and
 $2\omega$ fields. For a general value of $\theta$ the bichromatic
 driving force
 $f(t)$ breaks the time-reversal symmetry
 and a generalized parity \cite{flach}. Throughout we set $\hbar=1$, $T=1$ (hence all other parameters should be understood as
 scaled dimensionless variables).
 Typical values of these system parameters were well
 discussed in, e.g., Refs. \cite{holthaus0,weiss}.
 The value of $g$ is assumed to be tunable via the Feshbach
 resonance under an external magnetic field, though the  magnetic
 field is not explicitly included in the Hamiltonian.

For $f(t)=0$, Eqs. (\ref{Eq:dimer1}) and (\ref{Eq:dimer2}) have a
left-right permutation symmetry and reduce to the static double-well
model for self-trapping \cite{kladko,smerzi,ober}.  That is, for a
sufficiently large $g$, there exist two degenerate eigenstates with
a nonzero population imbalance $S\equiv N_1-N_2$.  For $f(t)\ne 0$,
in the linear limit the system (\ref{Eq:dimer1})-(\ref{Eq:dimer2})
possesses two Floquet states whose time evolution satisfies
$[\psi_{1}(T),\psi_{2}(T)]=[\phi_{1}(0),\phi_{2}(0)]\exp(-i\epsilon)$,
where $\phi_{1,2}(t+T)=\phi_{1,2}(t)$ and $\epsilon$ is the
quasienergy. The linear Floquet states can then be continuously
extended into the nonlinear domain \cite{njp}, thus defining
nonlinear Floquet states. Like in the stationary case, Floquet
states can also show population imbalance for large $g$ as a
manifestation of self-trapping.

To elucidate the properties of the nonlinear Floquet states we may
also map the mean-field dynamics of the two-mode system onto the
dynamics of a periodically driven classical system. Introducing the
variable $\varphi=\vartheta_{2}-\vartheta_{1}$, with
$\psi_{1}=|\psi_{1}|\exp{(i\vartheta_{1})}$ and
$\psi_{2}=|\psi_{2}|\exp{(i\vartheta_{2})}$, 
a classical effective Hamiltonian $H_{\text{eff}}$ exactly
describing the dynamics of $S$ and $\varphi$ is found to be
\cite{holthaus1} $H_{\text{eff}}=\frac{1}{2} g
S^{2}+C\sqrt{1-S^2}\cos(\varphi)+2 f(t)S$.
 The nonlinear Floquet states can then be mapped to the fixed points
of the Poincar\'e map of $H_{\text{eff}}$
\cite{holthaus0,holthaus1}.  

\begin{figure}
 \begin{center}
\begin{tabular}{lc}
\includegraphics[width=3.8cm,height=3.8cm]{Fig1a.eps}& \vspace{0.5cm}
\includegraphics[width=3.8cm,height=3.9cm]{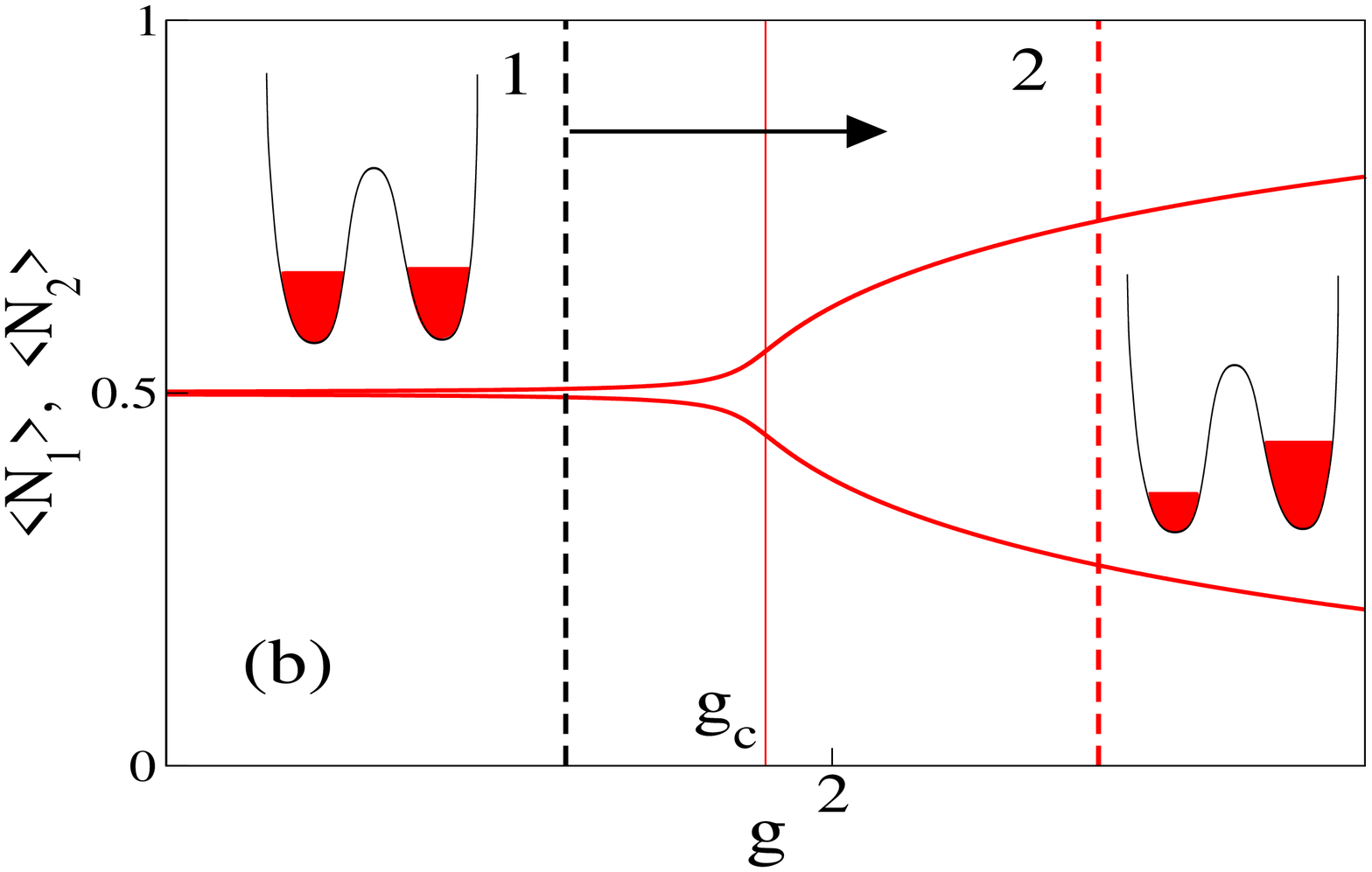}
\end{tabular}
\caption{(a) Quasienergy of nonlinear Floquet states vs the
nonlinearity strength $g$. Points $1$ and $2$ represent the initial
and final states in an adiabatic process (arrows indicate the
direction). The population imbalance of state $3$ on another branch
is almost the opposite of that for state $2$. (b) Populations for
each well averaged over a driving period as $g$ is adiabatically
increased. The dashed lines labeled by 1 or 2 correspond to the
states in (a). The thin solid line marks the threshold value $g_{c}$
for the onset of self-trapping for $\theta=-1.6$. Insets depict the
populations of atoms in the two wells for states 1 and 2. Other
parameters are $\omega=2\pi$, $f_{1}=f_{2}=1$, $C=1$. }
\label{Fig:bifurca}
\end{center}
\end{figure}

To examine how the nonlinear Floquet states may be adiabatically
correlated, we first study their behavior
 as the nonlinear parameter $g$ increases.  As we increases $g$,
 one of the two Floquet states can bifurcate into three nonlinear
Floquet states at a critical value $g_{c}$.  For the special cases
of $\theta= 0, \pi$, the system conserves a generalized parity (time
reversal together with a left-right permutation), we get a pitchfork
bifurcation with two
 degenerate states with exactly opposite nonzero population imbalance $S$. When the
 above symmetry is broken a saddle node [Fig. \ref{Fig:bifurca}] appears instead,
where two of the new states have very close quasi-energies with
almost opposite $S$;  and the third state corresponds to a saddle
point on the Poincar\'{e} map of $H_{\text{eff}}$.  In this latter
case the critical nonlinearity value for the bifurcation is a
function of $\theta$, denoted $g_{c}(\theta)$. The arrows in Fig.
\ref{Fig:bifurca} show that with an adiabatic increase in $g$ the
system will continuously  follow the upper bifurcation branch
corresponding to a nonlinear Floquet state with
increasing $S$. 

Taking the parameter $g$ as an example, how slowly the system
parameters should change in order to achieve
 necessary adiabaticity?     To answer this question
 we consider cases of linear ramping,
 where $g$ is made to change linearly in time,
between $g=g_{1}$ and $g=g_2$,
 at a rate of $\alpha$.
To get a rough estimate of how small the ramping rate $\alpha$
should be to ensure adiabaticity, we use the Landau-Zener tunneling
formula of Floquet states in the linear limit \cite{breuer}.
 For a representative case such as that shown in
 Fig.\ref{Fig:bifurca},
 we find that $\alpha \sim 10^{-4}$ suffices for an adiabatic process
 for which the Landau-Zener tunneling probability becomes negligible. To check that we take $g_{1}=1.8$ and $g_{2}=2.2$ and consider one
initial state such as the state $``1"$ shown  in  Fig.
\ref{Fig:bifurca} (preparing such an initial state requires the
tuning of $S$ and $\varphi$ \cite{ober}, followed by a fast turn-on
of the driving field). After the ramping we let the state evolve for
some time and then examine the time dependence of $N_{1}$ and
$N_{2}$. For $\alpha\le 10^{-4}$, the obtained state after the
ramping process is essentially identical with the adiabatic state
associated with $g=g_2$ (or state ``2" in Fig. 1). The periodic
oscillations in $N_1$ and $N_2$ also exactly match the frequency of
the driving field.

Significantly,  if the main interest is in the control of the
population imbalance $S$, then the averaged $S$ over many cycles of
the driving field is more relevant and hence it is unnecessary to
achieve full adiabaticity. Consider then cases with considerably
rapid ramping, e.g., $\alpha=10^{-2}$ (so the ramping only need
$\sim 10^2$ field periods). As shown in Fig. 2, for such a ramping
rate that is two orders of magnitude larger than before, almost the
same time-averaged $S$ is obtained. This can be further appreciated
using an alternative perspective afforded by the classical effective
Hamiltonian $H_{\text{eff}}$.  As seen from the right panel in Fig.
2, if $\alpha$ is much larger than $10^{-2}$, self-trapping breaks
down. The classical trajectory after the ramping process is found to
be on a chaotic layer around a separatrix structure, which
corresponds to the state in the lowest bifurcation branch.  This is
understandable because after a fast ramping process, the system has
not evolved much; so if the initial state is the fixed point for
$g=1.8$ located at $\varphi\approx-0.3$ and $S\approx 0$, it will
necessarily find itself in the neighborhood of the saddle point for
$g=2.2$ located also at $\varphi\approx -0.3$ and $S\approx 0$.

By contrast, if $\alpha\leq 10^{-2}$, then the classical trajectory
remains in the neighborhood of the fixed point.  Averaging out the
oscillations around the fixed point we obtain an averaged $S$ close
to that in the fully adiabatic case. Evidently then, manipulation of
population imbalance may be effectively achieved within a relatively
short time scale.

\begin{figure}
 \begin{center}
\begin{tabular}{lc}
\includegraphics[width=4.cm,height=4.3cm]{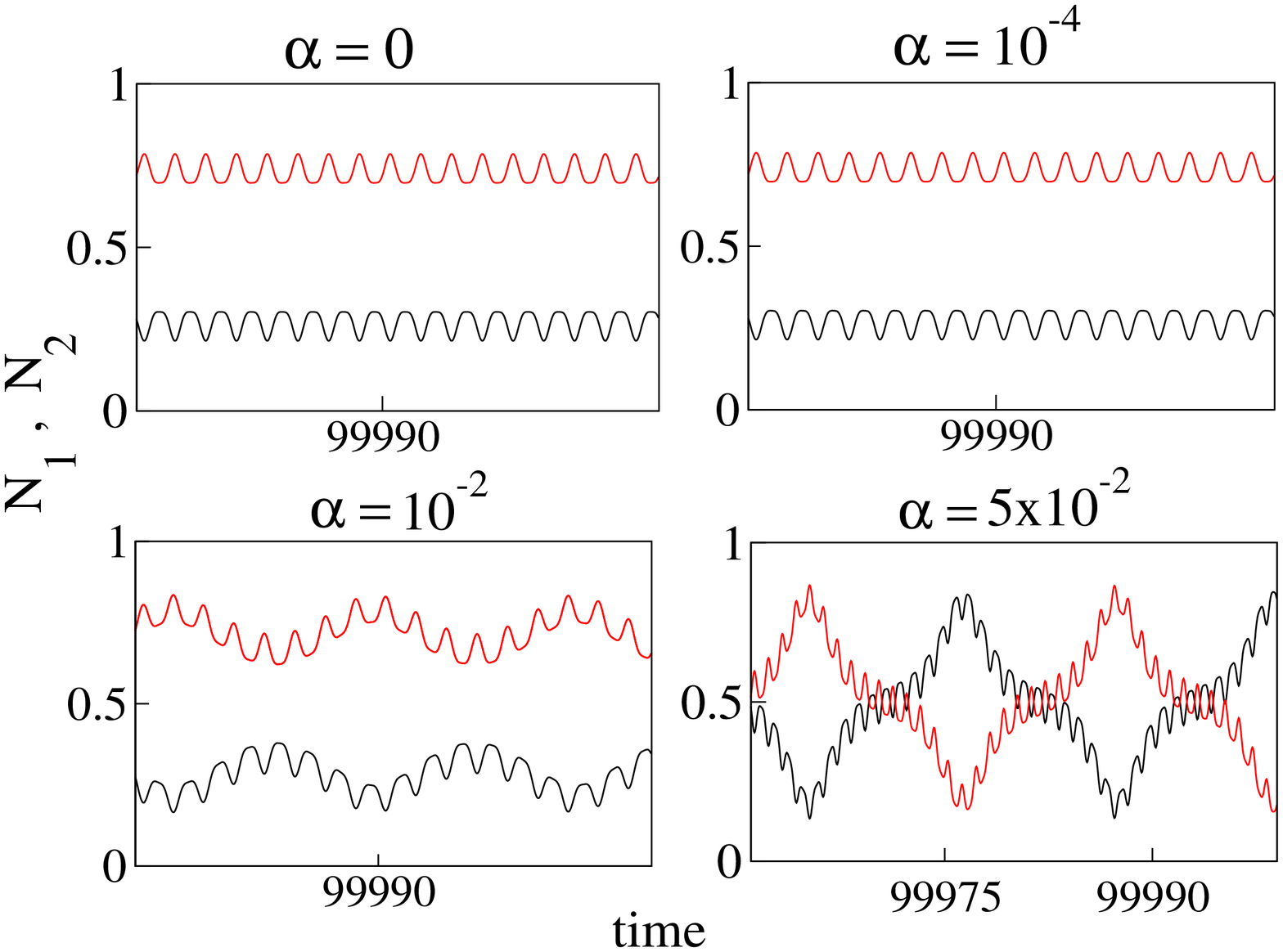}&
\includegraphics[width=3.8cm,height=4.cm]{Fig2b.eps}
\end{tabular}
\caption{Left panel: Evolution of the populations $N_{1}$ and
$N_{2}$ after ramping the nonlinearity parameter $g$ with different
rates $\alpha$. Right
  panel: The same dynamics as in the left panel
  plotted on the  $S$-$\varphi$ plane associated with $H_{\text{eff}}$ in
Eq.
  (4).
   Parameters are the same as in Fig. \ref{Fig:bifurca}. The case of $\alpha=0$
  represents
  the true adiabatic limit.}
\label{Fig:evol}
\end{center}
\end{figure}

We now ask if the population imbalance $S$ can be inverted. For that
purpose the parameter $\theta$ becomes crucial because it is
responsible for the symmetry breaking here.  However,  for strong
self-interaction $g>g_{c}$, a varying $\theta$  may have little
effect on a large $S$  due to the self-trapping effect. Further, a
path in the parameter space may not allow for adiabatic following if
certain loop structure of nonlinear Floquet states is encountered
(see, for example, case (c) of the lower panel of Fig.
\ref{Fig:Cross}). This motivates us to study how $S$ might be
inverted by tuning both $\theta$ and $g$.

Consider then a four-stage control cycle with state $2$ in Fig. 1
being the initial state. In stage I, the value of $g$ is ramped from
$g=2.2$ to $g=1.8$, a point where degeneracy of nonlinear Floquet
states no longer exists for any value of $\theta$.
 In stage II, $\theta$ is ramped from $-1.6$ to $1.6$.
 In stage III the value of
 $g$ is ramped back to $2.2$, and in stage IV $\theta$ is ramped back to
 $-1.6$. Stages III and IV are hence the opposite actions of stages
 I and II, but with a different order.
  As we ramp different parameters, we can have different ramping rates,
 e.g.,
 $\alpha_g$,
 $\alpha_{\theta}$. For simplicity, we set the two ramping rates to be the same.

\begin{figure}
\begin{center}
\begin{tabular}{lc}
\includegraphics[width=3.8cm,height=4.8cm]{Fig3a.eps}

\includegraphics[width=4.0cm,height=5.8cm]{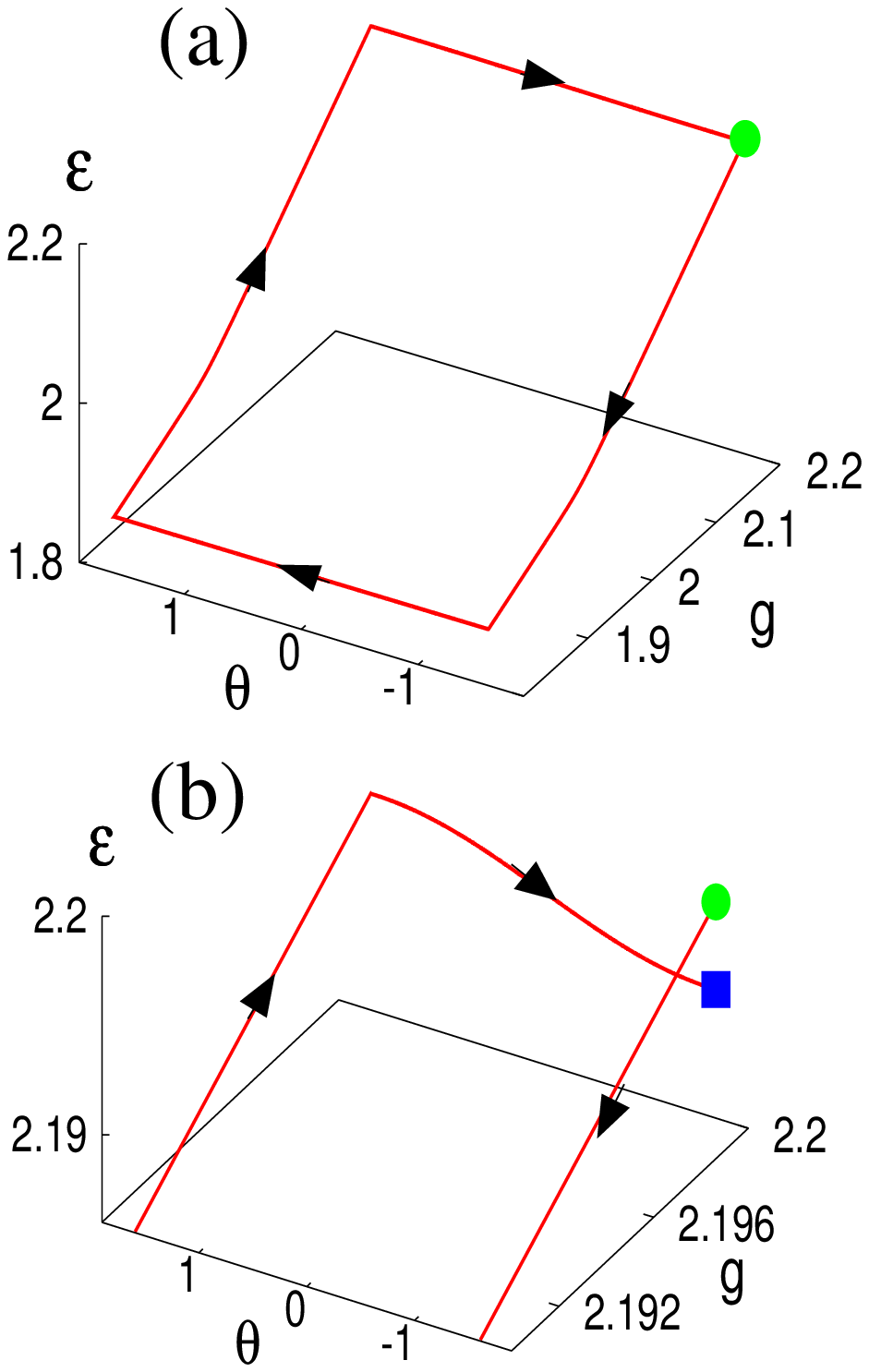}
\end{tabular}
\caption{Left panel: Population imbalance $S$ of the initial state
$2$ (bottom curve) and the final state (upper curves) after a
  four-stage ramping cycle.  $\alpha=10^{-4}$ for the upper solid curve and
  $\alpha=10^{-2}$ for the upper dotted-dashed curve.
  Inset:  the four-stage cycle with the
  dot being the starting point.
 Right panel: (a) Quasienergy along  the four-stage adiabatic cycle.
(b) Enlargement of the panel (a). The initial and final points are
depicted by a circle and square, respectively. Arrows indicate the
direction of the ramping process. }\label{Fig:surf}
\end{center}
\end{figure}

\begin{figure}
\begin{center}
\includegraphics[width=6.cm,height=5.2cm]{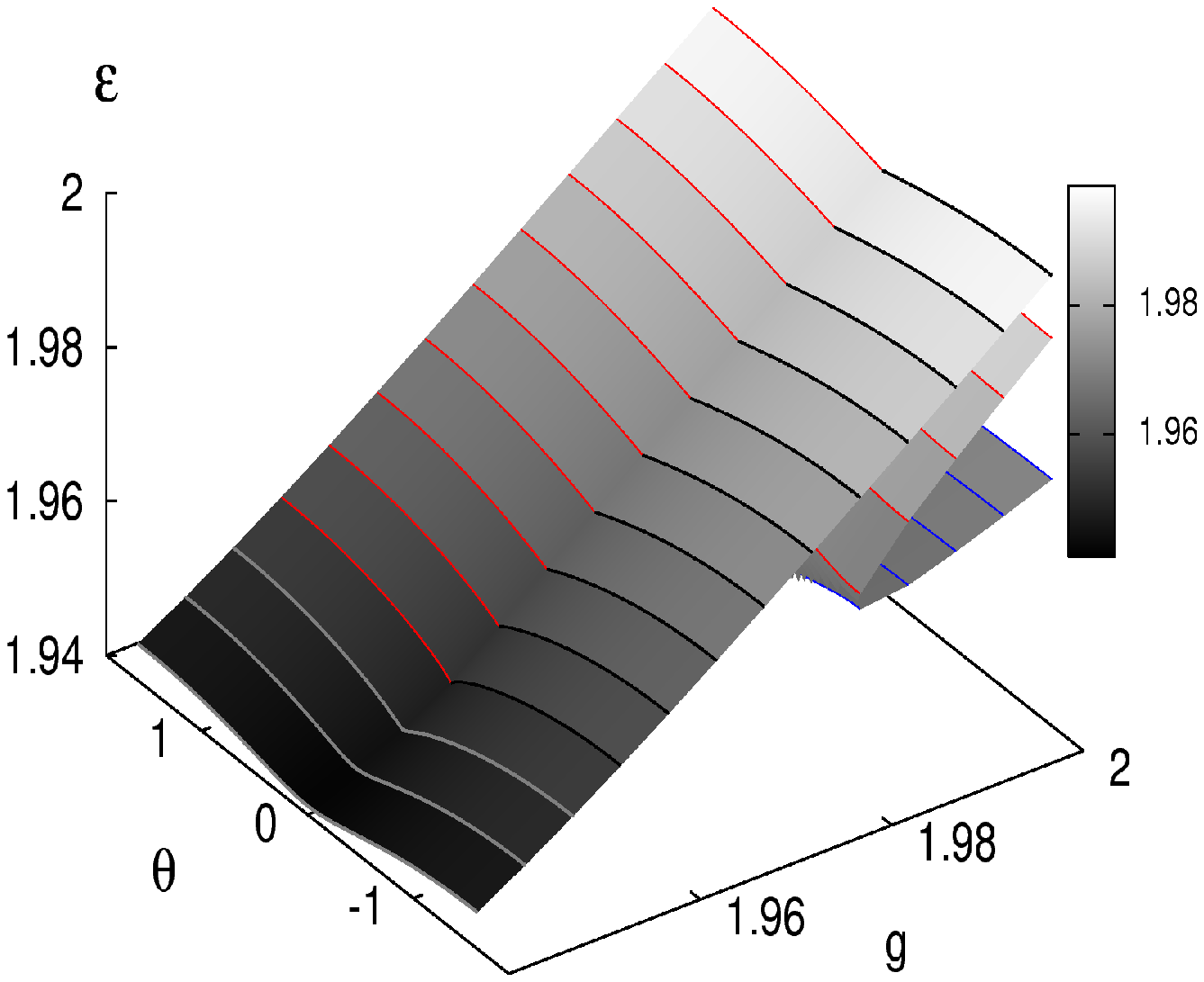}\\
\includegraphics[width=8.4cm,height=3cm]{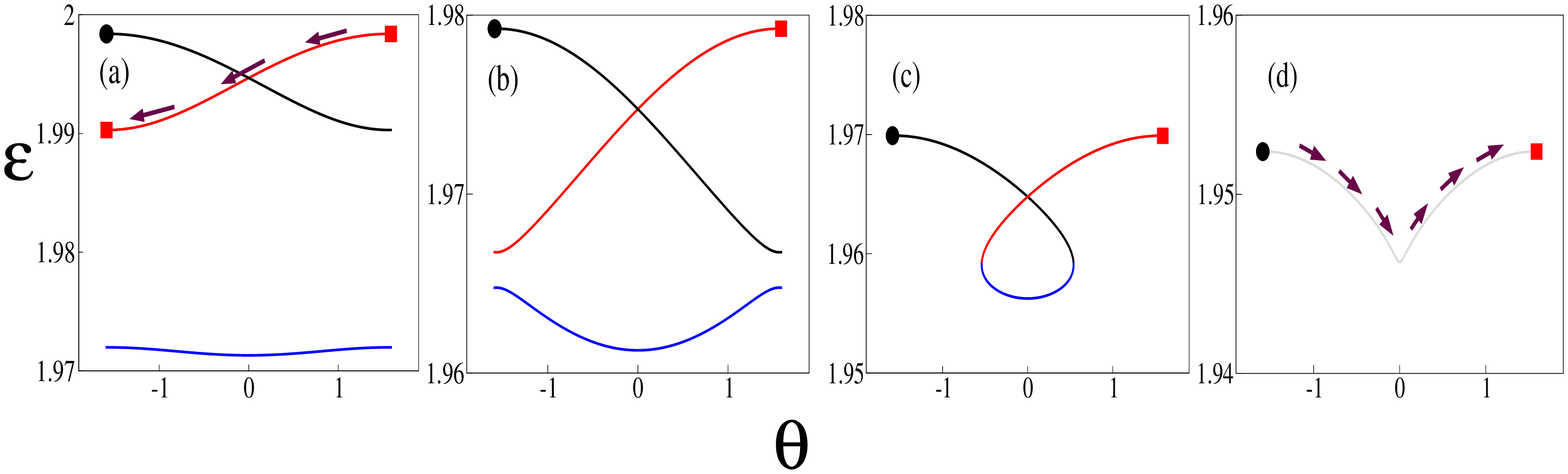}
\caption{Upper panel: Quasienergy surface vs the relative phase
parameter $\theta$ and the nonlinear parameter $g$. Lower panel:
Quasienergy $\epsilon$ vs $\theta$ for various fixed values of
$g$.(a) $g=2$; (b) $g=1.98$; (c) $g=1.97$; (d) $g=1.95$. The initial
state is depicted by a black dot in panel (a).
 In the first stage of the control cycle, $g$ decreases
 and the quasienergy state (black dot) changes as shown
 in the a-b-c-d sequence of the four panels. Arrows in
 panel (d) indicates the evolution of the quasi-energy in the
 second stage.
 This yields a state with almost opposite population imbalance
 depicted by the square.  In the third stage, in the order of d-c-b-a, the quasienergy (square)
 increases.  The evolution in the final stage is indicated by the
 arrows in panel (a). 
}\label{Fig:Cross}
\end{center}
\end{figure}

Figure \ref{Fig:surf} compares the time dependence of $S$ associated
with state 2 (bottom curve in the left panel) and that associated
with the final state  (upper curve with oscillations similar to
state 2) after the four-stage ramping cycle, with the linear ramp
rate in both $g$ and $\theta$ given by $\alpha=10^{-4}$. Clearly,
the resultant state after following the parameter cycle inverts $S$.
Moreover, its dynamics coincides with that of state 3 (see Fig. 1).
This indicates that after the four-stage control cycle, a nonlinear
Floquet state different from the initial one is reached. Next we
show in Fig. \ref{Fig:surf} the quasienergy along the control path.
At first glance, the quasienergy path seems to form a closed loop.
However, on a closer inspection of the last stage of the cycle (see
panel (b) on the right of Fig. 3), it is found that the quasienergy
of the final state (square) coincides with that of state 3 (see Fig.
1) and hence differs from that of the initial state (circle). This
again confirms that the system is placed on a different nonlinear
Floquet state after following the above ramping cycle.

To shed more light on the results we calculate the quasienergy
surfaces as a function of $\theta$
 and $g$  [Fig. \ref{Fig:Cross}]. For the sake of clarity, we only
 show the surface around the region where one quasienergy surface
 bifurcates into three surfaces.
 Above a critical strength of $g$, two quasi-energy surfaces can
 cross each other, forming a degeneracy line. Because this degeneracy line is along
 the $\theta=0$ direction (consistent with the fact that $f(t)$ with $\theta=0$
 conserves a generalized parity discussed above),
  the critical value $g$ for the surface crossing can be identified as
 $g_c(\theta=0)$.
  Note also that the degeneracy line does not present a problem for
  the system to
  continuously follow the ramping process,  because (i) $S$
  can only change continuously and (ii) the involved degenerate states however have opposite
  values of $S$.

Using the quasi-energy surfaces and its projection at different $g$
values,  we can now visualize the dynamics associated with the above
four-stage cycle.  In the first stage, the state slides down the
quasi-energy surface shown in Fig. 4, and at the same time three
nonlinear Floquet states are merged into one. At the end of the
first stage, the self-trapping phenomenon has disappeared and this
makes it possible to reverse $S$ by tuning $\theta$. The second step
does exactly this by allowing the symmetry-breaking driving force to
change $S$ from a very small, negative value to a very small, but
positive value. In the third stage, the $g$ value returns to its
original value, the system climbs up the quasi-energy surface, and
three nonlinear Floquet states emerge again as a result of
bifurcation. During this stage the inverted $S$ is then magnified in
the same mechanism as shown in Fig. 2. In the last stage when
$\theta$ is ramped back, the system can maintain its population
imbalance due to self-trapping. In the meantime the system crosses
the degeneracy line, returns all system parameters to their initial
values, and finally ends up in a different Floquet state.  This
analysis also makes it clear that by considering a reversed closed
path one can also change $S$ from positive to negative.

One may regard the above ``anholonomy-like" behavior as a
consequence of adiabatic following during the entire ramping
process, insofar as the system always follows the control cycle and
never populates two nonlinear Floquet states. However, due to the
nonlinearity-induced degeneracy line along $\theta=0$, it is better
to regard the evolution during stage IV as a {\it diabatic} process.
That is, during stage IV the change of the parameters is always
``too fast" as compared with the zero spacing at the degeneracy
point. Indeed, in a fully quantum treatment without the mean-field
approximation, the quantum levels, though nondegenerate, will be
highly clustered around the mean-field degeneracy point. Therefore
in a fully quantum picture adiabatic following breaks down during
stage IV. In either perspective, the net result is the same: the
system finds itself in a different state after following a ramping
cycle in the parameter space.

We next examine whether we can achieve an inversion of $S$ if the
system parameters are tuned at a much faster rate? Significantly, as
already suggested by the results in Fig. 2,  the adiabatic following
approach here still works even if we ramp the parameters $g$ and
$\theta$ at a considerable rate. Indeed, the result for a ramping
rate two orders of magnitude larger ($\alpha=10^{-2}$) is also shown
in Fig. 3 (dotted-dashed curve on the left panel). In this case, as
indicated by the beating patterns of the dynamics, the final state
is not on one single Floquet state, but upon a time average we still
achieve an inverted $S$ close to that obtained from the previous
slow ramping case. We have also checked many other clockwise
(counter-clockwise) closed paths in the $g-\theta$ plane of Fig. 4,
finding that they can yield similar control from state 2 to state 3
(from state 3 to state 2) if they enclose the critical point
$g=g_c(\theta=0); \theta=0$. Thus, we conclude that, given the
topology of quasi-energy surfaces, especially the degeneracy and
bifurcation points, navigation of nonlinear Floquet states provides
a useful control scenario. This also extends an early control
scenario based on the navigation of linear Floquet states
\cite{luisEPL}.  Applying our approach to the cases with nonlinear
Floquet states localized in momentum \cite{njp}, it should be
possible to invert ratchet transport in many-body systems without a
biased field.

In conclusion, by demonstrating the control of the population
imbalance associated with a BEC in a double-well potential, we have
shown that the navigation of nonlinear Floquet states associated
with symmetry-breaking driving fields can offer effective and robust
control over many-body systems on the mean-field level.

{\bf Acknowledgments}: We are grateful to Dr. Sergej Flach for
interesting discussions and critical comments on early versions of
this manuscript.
 One of the authors (J.G.) is supported by the start-up funding
(WBS grant No. R-144-050-193-133/101) and the NUS ``YIA'' funding
(WBS grant No. R-144-000-195-123). J.G. also thanks Zhang Qi for
helpful discussions.

\end{document}